\documentclass[american,twocolumn, showpacs,prl,floatfix]{revtex4}
\usepackage[T1]{fontenc}
\usepackage[latin1]{inputenc}
\usepackage{graphicx}
\usepackage{amssymb}
\usepackage{slashed}
\usepackage{color}
\usepackage{babel}

\begin{document}

\title{Yoctosecond photon pulses from quark-gluon plasmas}

\date{October 15, 2009}

\author{Andreas \surname{Ipp}}
\email{andreas.ipp@mpi-hd.mpg.de}

\author{Christoph H.~\surname{Keitel}}

\author{J\"org \surname{Evers}}
\email{joerg.evers@mpi-hd.mpg.de}

\affiliation{Max Planck Institute for Nuclear Physics, Saupfercheckweg 1, D-69117
Heidelberg, Germany}

\begin{abstract}
Present ultra-fast laser optics is at the frontier between atto- and
zeptosecond photon pulses, giving rise to unprecedented applications.
We show that high-energetic photon pulses down to the yoctosecond
timescale can be produced in heavy ion collisions. We focus on 
photons produced during the initial phase of the expanding
quark-gluon plasma. We study how the time evolution and properties
of the plasma may influence the duration and shape of the photon pulse.
Prospects for achieving double peak structures suitable for pump-probe
experiments at the yoctosecond timescale are discussed.
\end{abstract}


\maketitle
Recent advances in ultrafast laser optics aim at the creation of ultra-short pulses beyond the visible spectral range. 
Shorter pulses are desirable, since they provide better temporal resolution, while higher photon energy gives rise to better spatial resolution. High-order harmonics of femtosecond laser radiation have been shown to be sources of trains of attosecond 
extreme-ultraviolet pulses~\citep{Papadogiannis1999,Paul2001} that can be used to produce single attosecond
soft
X-ray~pulses~\citep{Hentschel:2001,Silberberg2001}. For example, such single attosecond X-ray bursts~\citep{baeva07} have applications in molecular imaging~\citep{niikura02,Lein2005},
quantum control~\citep{Rabitz2000,Silberberg2004}, or Raman spectroscopy~\citep{Dudovich2002}. By introducing a controlled delay between two such peaks, the dynamics of electron systems could be studied using pump-probe techniques \citep{Silberberg2001}. 
Such techniques also allow for the direct time resolution of 
many-body dynamics, like the observation of
the dressing process of charged particles~\citep{Huber:2001}.
In view of the obvious desire for even shorter pulses with higher photon energies, it is reasonable to search for alternative production methods. In this spirit, it has been suggested that zeptosecond pulses could be created by 
focusing intense laser pulses on subwavelength-size structures~\citep{PhysRevLett.88.074801}, by the reflection of a relativistically intense femtosecond laser pulse from the oscillating boundary of an overdense plasma~\citep{gordienko04}, or via nonlinear Thomson backscattering~\citep{lan:066501}.

Among the shortest possible time scales that are available experimentally are those obtained through high-energy collisions. Particularly interesting in this context are heavy ion collisions that can produce a quark-gluon plasma (QGP), because they expose a complex system with rich internal dynamics that lives on a very short timescale. Heavy-ion collisions at the Relativistic Heavy Ion Collider (RHIC) and soon at the CERN LHC produce this new state of matter up to the size of a nucleus ($\sim15\,\mbox{fm}$) for a duration of a few tens of yoctoseconds ($1\,\mbox{ys}=10^{-24}\,\mbox{s}\approx0.3\,\mbox{fm/c}$). In such a collision, the plasma is produced initially in a very anisotropic state, and reaches a hydrodynamic evolution through internal interactions only after some thermalization time $\tau_{\rm therm}$ and isotropization time $\tau_{\rm iso}$. The observed particle spectra turned out to agree well with ideal hydrodynamical model predictions~\citep{Huovinen:2001cy,Hirano:2002ds,Tannenbaum:2006ch}, which led to the assumption that 
isotropization times are as low as $\tau_{\rm iso}\approx1\,\mbox{ys}(0.3\,\mbox{fm/c})$.
However, it has been pointed out recently that viscous hydrodynamic models are still consistent with RHIC data if isotropization times as large as $\tau_{\rm {iso}}\approx7\,\mbox{ys}(2\,\mbox{fm/c})$ are assumed \citep{Luzum:2008cw}, even if the expansion before isotropization is assumed to be collisionless ({}``free streaming''). Besiders a plethora of particles that are
created in such collisions, also high-energetic photons are produced and detected experimentally \citep{Adams:2004fm,Adler:2005ig,ipp}. 
So far, only time-integrated quantities
have been considered, like the extraction of the spatiotemporal extent of the emitting source through photon pair correlations~\citep{Marques:1994}
and the suggestion to measure the isotropization time $\tau_{\rm {iso}}$ using the direct photon yield~\citep{Schenke:2006yp,Bhattacharya:2008up}.

In this Letter, we study the time-resolved production of high-energetic ultra-short photon pulses in the QGP. We demonstrate that the emission envelope depends strongly on the internal dynamics of the QGP.  Under certain conditions, a double peak structure in the emission envelope can be observed, which could be the first source for pump-probe experiments at the yoctosecond timescale. We find that the delay between the peaks is directly related to the isotropization time, and the relative height between the peaks can be shaped by varying photon energy and emission angle.  Such pulses could be utilized, for example, to resolve dynamics on the nuclear timescale such as that of baryon resonances~\citep{Dugger:2007bt}. As an alternative interpretation of our results, a time-resolved study of the emitted photons could provide a window to the internal QGP dynamics throughout its expansion.

\begin{figure}[t]
\centering
\includegraphics[width=7.5cm]{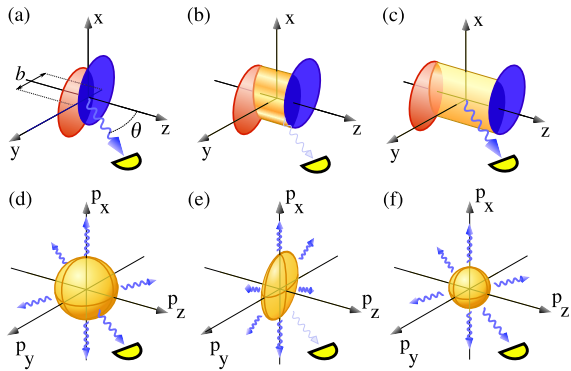}
\caption{\label{fig-system}(color online) 
Early stages of a high-energy collision, involving pre-equilibrium (first two columns) and equilibrated QGP phases (last column).
Parts (a)-(c) show three snapshots in time in position space. Shown are the two relativistically contracted colliding ions that create the quark-gluon plasma in the overlap region. Curly arrows denote photon emission and semicircles the detectors. Parts (d)-(f) are corresponding pictorial representations of the plasma in momentum space. In an intermediate stage
of the pre-equilibrium phase,
the momentum distribution is anisotropic, resulting in a change in the angular photon emission pattern that can give rise to double-peaked photon pulses.}
\end{figure}

The QGP is formed in a collision of two heavy ions as illustrated in Fig.~\ref{fig-system}(a-c). We study the emission of direct photons from the expanding QGP~\citep{Reygers:2006qb,Turbide:2003si}.  The energy spectrum of such photons extends to the GeV range, and the upper limit for the temporal duration of the GeV photon pulse is given by the expansion dynamics of the QGP, which leads to yoctosecond pulses. Regarding the shape of the photon pulse, we find that the angle-resolved photon emission rate strongly depends on the internal state of the plasma, characterized by the momentum distribution of the plasma constituents, see Fig.~\ref{fig-system}(d)-\ref{fig-system}(f). For an intermediate time after the collision, a momentum anisotropy occurs 
due to the longitudinal expansion of the plasma, 
which leads to a preferential emission of photons perpendicular to the beam axis $z$. 
A photon detector placed towards the beam axis would therefore measure a time-dependent photon flux.
Based on a recent model~\citep{Mauricio:2007vz,Martinez:2008di} for the internal plasma dynamics, we find that this mechanism can give rise to double-peaked pulse envelopes that could be utilized for pump-probe experiments.

We base our model on the one-dimensional expansion model by Bjorken \citep{Bjorken:1982qr}. This assumes boost-invariant evolution of the quark-gluon plasma in a central region of the collision. The detector is placed away from the beam axis by an angle $\theta$
within the reaction plane.
Since we want to study hard photons produced in the GeV energy range,
we restrict ourselves to the production of direct photons 
emitted from the pre-equilibrium and
equilibrated phases of the QGP. We will not consider 
photons produced in later stages of the collision where
on average lower energy photons are emitted. 
For the same reason, we will omit higher order soft scattering
processes like bremsstrahlung or inelastic pair annihilation,
whose contributions are again more important at lower energies~\citep{ipp,Schenke:2006yp}.
The leading contribution to the photon production rate $R$ originates from quark-gluon Compton scattering and quark-antiquark annihilation.
For anisotropic momentum distributions, the photon production rate $R$ 
has to be calculated numerically~\citep{ipp}.
This rate, obtained in the form $E {d^{3}R}/ {d^{3}k}$, depends on the temperature $T$ of the medium, the photon energy $E$ and momentum $k$, the fine structure constant $\alpha$, and the corresponding quantity for the strong force $\alpha_{s}$ (we use $\hbar=c=k_{\rm B}=1$).
The rate further depends on the anisotropy, which is described by a parameter 
$\xi=\left\langle p_{T}^{2}\right\rangle /\left(2\left\langle p_{L}^{2}\right\rangle \right)-1$
that relates the mean longitudinal and transverse momenta 
$p_{L}$ and $p_{T}$~\citep{Schenke:2006yp,Mauricio:2007vz,Martinez:2008di}.
To integrate this rate over time, we use a recent model to describe the time evolution of the pre-equilibrium and equilibrated QGP~\citep{Mauricio:2007vz,Martinez:2008di}.
This model specifies the time evolution for the energy density $\mathcal{E}=\mathcal{E}(\tau)$, for the hard scale $p_{{\rm hard}}=p_{\rm hard}(\tau)$ (which corresponds to $T$ in the isotropic case), and for the anisotropy
parameter $\xi=\xi(\tau)$ as a function of the proper time $\tau$. These quantities scale with different powers depending on the QGP expansion dynamics. For example, the time evolution of $\xi$ can be written as $\xi(\tau)=(\tau/\tau_{0})^{\delta}-1$. In the free streaming phase, $\delta=2$, while in the ideal hydrodynamic phase $\delta=0$.  
The model in Ref.~\citep{Martinez:2008di} essentially introduces a smeared step-function for the exponent $\delta=\delta(\tau)$ to interpolate between $\delta=2$ 
at early times
$\tau\ll\tau_{\rm {iso}}$ and $\delta=0$ at late times $\tau\gg\tau_{\rm {iso}}$, where the duration of the transition is controlled by $\tau_{\rm {iso}}/\gamma$ with dimensionless parameter $\gamma$. 
In this way, both the pre-equilibrium phase and the equilibrated QGP phase of
the expansion can be described by a single model.
In the model, thermalization and isotropization happen concurrently, $\tau_{\rm {therm}}=\tau_{\rm {iso}}$.

In the radial direction, we use a straightforward generalization of the temperature dependence $T(r,\tau_{0})=T_{0}[2(1-r^{2}/R_{T}^{2})]^{1/4}$~\citep{Turbide:2005fk,Fries:2002kt} for non-central collisions that is valid in the overlap region of the colliding nuclei. As in Ref.~\citep{Martinez:2008di}, we neglect the expansion of the QGP into transverse directions, since it is small in the initial stage throughout which the high-energetic photons are emitted. $T_{0}$ is the initial temperature, $r$ is the distance from the center in radial direction, and $R_{T}$ is the transverse radius of the nucleus. 
The number of photons of frequency $\omega_{d}$ that arrive at the detector at time $t_{d}$ in the laboratory system can be obtained by integrating the photon rate 
along all possible light paths. The world line of the photons 
is parametrized as 
$\mathbf{z}(t;t_{d},x_{k},y_{k})=x_{k}\mathbf{\hat{k}}_{1}^{\perp}+y_{k}\mathbf{\hat{k}}_{2}^{\perp}+\mathbf{\hat{k}}\left(t-t_{d}+d\right)$,
where $\mathbf{\hat{k}}$ is the spatial direction of the light wave vector $k=(\omega,\mathbf{k})$, $\mathbf{\hat{k}}_{1}^{\perp}$ and $\mathbf{\hat{k}}_{2}^{\perp}$ are two direction vectors orthogonal to $\mathbf{k}$ and to each other, and $d$ is the distance to the detector. 
The signal $f_{d}$ that arrives at the detector at time $t_{d}$ is given by integrating the photon emission
rate along the photon world line
$f_{d}(k,t_{d},x_{k},y_{k})=\int_{0}^{t_{d}}dt\, f(k,t,\mathbf{z} (t;t_{d},x_{k},y_{k}))$
where $f(k,t,\mathbf{z})$ is the Lorentz transformed photon emission rate 
$E\, d^{3}R/d^{3}k$ inside the QGP, and vanishing outside.
One finally obtains the photon rate per unit time and frequency interval with $A_{d}$ the detector area,
\begin{equation}
\label{eq:detectorPhotonRate3d}
\frac{d^{2}N(\omega_{d},t_{d})}{dt_{d}d\omega_{d}}=\frac{A_{d}}{8\pi d^{2}}\int dx_{k}\, dy_{k}\,\omega_{d}f_{d}(\omega_{d},t_{d},x_{k},y_{k})\,.
\end{equation}
%

%
\begin{figure}
\centering
\includegraphics[width=0.95\columnwidth]{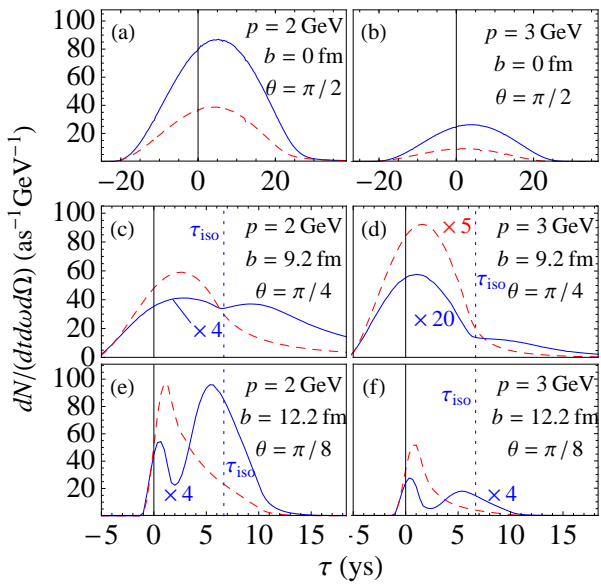}
\caption{\label{fig:result}(color online) Photon emission rate as a function of detector time $\tau$.
Solid blue lines show a large isotropization time $\tau_{\rm {iso}}=6.7\,\mbox{ys }=2\,\mbox{fm/c}$ with $\delta=2$ (free streaming model) while dashed red lines correspond to a short isotropization time $\tau_{\rm {iso}}=\tau_{0}=0.3\,\mbox{ys}=0.088\mbox{\, fm/c}$. 
Parts (a) and (b) display emission at midrapidity ($\theta=\pi/2$) for a central collision with impact parameter $b=0$.
Parts (c)-(f) show double-peaked 
photon pulses obtained for $b=9.2$~fm or $12.2$~fm, and the vertical dotted line indicates the position of the larger $\tau_{\rm {iso}}=6.7$~ys. 
In parts (c) and (d), the detector direction is $\theta=\pi/4$, in (e) and (f), it is $\theta=\pi/8$.
}
\end{figure}

For the concrete calculation we followed the parameters as given in \citep{Martinez:2008di}. 
We show results for LHC energies, 
where the initial conditions are given by a formation time $\tau_{0}=0.3\,\mbox{ys}\,(=0.088\mbox{\, fm/c})$, 
the initial temperature $T_{0}=845$ MeV, 
and transverse radius of the lead ion $R_{T}=7.1$ fm. 
The critical temperature, where the QGP ceases to exist, is taken as $T_{C}=160$ MeV. 
We study the free-streaming model ($\delta=2$) with parameter $\gamma=2$. 
As in Ref.~\citep{Martinez:2008di}, the isotropization time is varied in the range $\tau_{{\rm {iso}}}=\tau_{0}$ to $\tau_{{\rm {iso}}}=6.7\,\mbox{ys}\,(=2\,\mbox{fm/c})$.
Both possibilities are not yet ruled out by RHIC data. We use the model with fixed final multiplicity, where the initial conditions are adjusted as a function of $\tau_{\rm {iso}}$ in order to result in the same final entropy as for $\tau_{\rm {iso}}=\tau_{0}$.

Figures~\ref{fig:result}(a) and \ref{fig:result}(b) show the typical time evolution of the photon emission rate for a central collision with emission angle orthogonal to the beam axis ($\theta=\pi/2$).
We present results at 2 GeV (3 GeV) energy, 
where the photon production from the QGP at midrapidity
is 3 to 4 times as large as the production from the initial collisions, 
and roughly 6 times (50 times) as large as the 
production from the hadron gas~\citep{Turbide:2003si}.
 The origin of the abscissa is the time when a photon emitted from the center of the collision arrives at the detector. Photons arriving earlier originate from a part of the QGP that is closer to the detector. The pulse shape is mainly determined by the geometry of the nucleus with radius $7.1\,\mbox{fm}$. 

Internal QGP dynamics occurs on a timescale of a few fm/c. Any structure of this order is washed out simply by the time for light to cross the size of the QGP. We apply two strategies to overcome this limit. First, we reduce the physical extent of the QGP by considering non-central collisions with impact parameter $b$. Second, an optimization of the detection angle minimizes the traveling time through the plasma. In forward direction, the initial shape of the QGP is Lorentz-contracted, hence light travels through the initial shape quickly. This is partially spoiled due to the QGP expansion in the same direction. Thus intermediate emission angles are most promising for which the QGP appears partly Lorentz contracted but does not expand towards the detector.

Figures~\ref{fig:result}(c)-\ref{fig:result}(f) show the photon emission in a direction away from midrapidity $\theta\neq\pi/2$ for a non-central collision. The impact parameter $b=9.2$~fm ($b=12.2$~fm) corresponds to an overlap region of the size $5\,\mbox{fm}\times10.8\,\mbox{fm}$ ($2\,\mbox{fm}\times7.3\,\mbox{fm}$).
 Here, a striking new double-peak structure appears. Roughly speaking, the minimum between the peaks corresponds to the maximum of the anisotropy parameter $\xi(\tau)$, which grows with exponent $\delta=2$ for $\tau<\tau_{\rm {iso}}$, and vanishes for $\tau>\tau_{\rm {iso}}$. This is because the photon emission
rate is suppressed for larger values of $\xi$ and smaller values of $\theta$~\citep{Schenke:2006yp}. The distance between the two peaks is therefore approximately governed by $\tau_{\rm {iso}}$, indicated by the dotted line in Figs.~\ref{fig:result}(c)-\ref{fig:result}(f). 

The dynamics shown in Fig.~\ref{fig:result} requires a more detailed interpretation. 
In Figs.~\ref{fig:result}(c)-\ref{fig:result}(f), the first peak  corresponds to photons emitted from the blue-shifted approaching part of the QGP, 
while the second peak corresponds to 
a slightly red-shifted and time-dilated receding tail of the plasma. 
The relative size of the first and second peak is controlled both by the emission angle and the photon energy. 
At increasing emission angles, the second peak is suppressed,
because the phase space for photons emitted from the receding part of the plasma
is constricted.
For smaller impact parameters or larger QGP sizes, the second peak overlaps with the first, thereby washing out this structure.
Also for a short isotropization time $\tau_{\rm {iso}}=\tau_{0}$ (dashed lines) the separation into two peaks does not occur. Therefore this effect is very sensitive to the internal dynamics of the QGP.

Our results are based on a number of model assumptions. For example, we 
omitted
the transverse expansion, which would become important at later times. It could either enhance the double peak structure through a further separation of red- and blue-shifted parts of the QGP, or reduce the effect through a prolonged photon passage through the QGP. These modifications may cause qualitative changes to the photon emission envelope, but not to the yoctosecond timescale.
In an actual experiment there is a background of photons from different sources~\citep{Turbide:2005fk}. These include photons produced by a jet passing through the QGP~\citep{Fries:2002kt}, and could dominate the effect that is expected from the QGP alone. Since these photons are produced on a similar yoctosecond timescale, they can be expected to have the positive effect of enhancing the photon rate of the pulse. But at the same time they would constitute a background for the pulse shape determination. Other background photons are produced at different time scales, e.g., due to decay of pions produced in the hadronization of the QGP,
therefore they would not modify a time structure on the yoctosecond timescale.
Photons produced from the initial collisions (pQCD photons) can be of comparable size,
but they would only enhance the first peak of the double peaks depicted in Fig.~\ref{fig:result}(c)-\ref{fig:result}(f).
A better theoretical understanding of the QGP dynamics would be necessary for a more quantitative prediction
of the expected pulse structure.

We now turn to an estimate of the photon production rate. In the GeV energy range, these are of the order of a few photons per collision~\citep{Turbide:2005fk}. Note that a single GeV photon pulse of 10~ys duration corresponds to a pulse energy of only about $100$~pJ, but to a power of $10$~TW. 
A single photon per pulse is in principle sufficient
to reconstruct a non-trivial pulse shape~\citep{Keller:2004}.
Decreasing the photon energy would enhance the photon yield, but would also 
increase the number of unwanted background photons.
Increasing the collision energy would further enhance the number of photons produced, 
and could also increase the relative importance of the contribution of thermal photons compared to other kinds of photons~\citep{Turbide:2005fk}. 

Concluding, we have studied the time evolution of the photon emission from the QGP. Since the QGP only exists on the yoctosecond timescale, it naturally emits photons at this timescale. We have shown that the emission envelope can be influenced by the geometry, emission angle, and internal dynamics like the isotropization time of the expanding QGP. For a particular parameter range, 
that is non-central collisions, large isotropization time, and an emission angle close to forward direction, 
a double-peak structure has been found within our model.  Thus, pump-probe experiments at the GeV energy scale could be envisioned. 
New detection schemes would be required, and it should be explored whether
tools and ideas from attosecond metrology~\citep{Hentschel:2001,Drescher:2001} can be scaled to zepto- or 
yoctosecond duration, utilizing physical processes in the GeV energy range, like electron-positron pair creation.
Alternatively, determining the photon emission shape experimentally would give direct access to dynamic properties of the QGP, like its isotropization time.

\acknowledgements
We thank K.~Z.~Hatsagortsyan and M.~Strickland for helpful discussions.


\end{document}